\documentclass[12pt]{article} 
\usepackage{epsfig}
\usepackage{pifont,color} 
\usepackage{amsmath,amssymb}
\usepackage{citesort}

\usepackage{axofigs}

\def \cc  {\overline}
\def \lm  {\lambda}
\def \To  {\Rightarrow}
\def \bra {\langle}
\def \ket {\rangle}
\def \hws {| h, \cc{h} \ket}
\def \lf  {\left (}
\def \rt  {\right )}
\def \pl  {\partial}
\def \scri {\ensuremath{\mathcal{I}}}
\def \scriplus {\ensuremath{\mathcal{I}^+}}
\def \scriminus {\ensuremath{\mathcal{I}^-}}
\def \ZZ  {\mathbb{Z}_2}
\newcommand{\dimmin}[2]{$#1\!-\!#2$-dimensional}
\newcommand{\dimplus}[2]{$#1\!+\!#2$-dimensional}

\begin{document}

\begin{titlepage}

\begin{center}

{\hbox to \hsize{\hfill hep-th/0209120}}
{\hbox to \hsize{\hfill CU-TP-1065}}
{\hbox to \hsize{\hfill SPIN-2002/24}}
{\hbox to \hsize{\hfill PUPT-2045}}

\bigskip

\vspace{6\baselineskip}

{\Large \bf \mbox{Elliptic de Sitter Space:} $dS/\ZZ$}

\bigskip
\bigskip
\bigskip

{\large \sc Maulik~Parikh,\footnote{\tt mkp@phys.columbia.edu}
  Ivo~Savonije,\footnote{ \tt savonije@phys.uu.nl} {\rm and}
  Erik~Verlinde\footnote{\tt erikv@feynman.princeton.edu}\\[1cm]
}

{\it $^1$ Department of Physics, Columbia University, New York, NY
  10027}\\[3mm]
{\it $^2$ Spinoza Institute, University of Utrecht, Utrecht, The
  Netherlands}\\[3mm]
{\it $^3$ Physics Department, Princeton University, Princeton, NJ
  08544}\\[3mm]

\vspace*{1.5cm}

\large{
{\bf Abstract}\\
}
\end{center}
\noindent
We propose that for every event in de Sitter space, there is a
CPT-conjugate event at its antipode. Such an ``elliptic''
$\ZZ$-identification of de Sitter space provides a concrete
realization of observer complementarity: every observer has complete
information. It is possible to define the analog of an S-matrix for
quantum gravity in elliptic de Sitter space that is measurable by all
observers. In a holographic description, S-matrix elements may be
represented by correlation functions of a dual (conformal field)
theory that lives on the single boundary sphere. S-matrix elements are
de Sitter-invariant, but have different interpretations for different
observers. We argue that Hilbert states do not necessarily form
representations of the full de Sitter group, but just of the subgroup
of rotations. As a result, the Hilbert space can be finite-dimensional
and still have positive norm. We also discuss the elliptic
interpretation of de Sitter space in the context of type IIB* string
theory.
                          
\end{titlepage}

\newpage

\setcounter{page}{2}
\setcounter{footnote}{0}

\tableofcontents

\bigskip
\bigskip

\section{Introduction}
\label{sec:introduction}

In a monograph first published in 1956, Schr\"odinger
\cite{Schroedinger} describes a troubling consequence of the
exponential expansion of space in a de Sitter universe, namely that
different observers would be swept out of each other's event horizons:
\begin{quote}
...it does seem rather odd that two or more observers, even such as
``sat on the same school bench'' in the remote past, should in future,
when they have ``followed different paths in life,'' experience
different worlds, so that eventually certain parts of the experienced
world of one of them should remain {\em by principle} inaccessible to
the other and vice versa.
\end{quote}
The separation of spacetime into causally inaccessible regions is not
just unaesthetic, but conceptually problematic. It suggests, for
instance, that pure states could evolve into mixed states, as
degrees of freedom disappear across the horizon. For an observer in de
Sitter space this would manifest itself as quantum decoherence and a
loss of information.

Similar issues arose in the study of the information loss problem for
black holes. Gedankenexperiments in that context essentially led to
the conclusion that unitarity could be preserved for all observers if
one allowed for a duplication of information on either side of the
horizon. According to this ``principle of black hole complementarity,''
\cite{stu,thooft,compl} the freely-falling observer and the external
observer would both be able to perform quantum mechanics experiments
without any loss of coherence, but their interpretation of the physics
would be quite different.

The arguments that lead to black hole complementarity can also be
applied to other types of event horizons, in particular to
cosmological event horizons. A better name therefore would be
``observer complementarity.'' In its strongest form it postulates that
each observer has complete information, and can in principle describe
everything that happens within his cosmological horizon using pure
states. This information may appear to different observers in
different -- complementary -- guises: one observer may pass smoothly
through the horizon, whereas another observer may see there a source
of hot radiation. Although these drastically different realities may
seem to be inconsistent, it is important to recognize that paradoxes
arise only when one takes the unphysical perspective of a global
super-observer.

The question now is, is there a way to implement observer
complementarity in de Sitter space? There is, as was already noted by
Schr\"odinger. In his ``elliptic interpretation''\footnote{The term
``elliptic'' refers to the fact that identified points are related by
elliptic, i.e. spacelike, generators, as distinct from hyperbolic
(timelike) or parabolic (null) generators.}  of de Sitter space,
Schr\"odinger proposed a simple $\ZZ$ identification of spacetime by
declaring antipodes to represent the same event. Schr\"odinger's
motivation was indeed to give all observers complete information about
all events, and thus in a way he argued already in 1956 in favor of
observer complementarity. In this paper, we consider the consequences
of the elliptic interpretation. We find that elliptic de Sitter space
has some rather remarkable properties. Indeed, not only does it lead
to a concrete realization of observer complementarity, it also
improves the nature of many of the severe theoretical challenges that
de Sitter space presents. The main aim of this paper therefore is to
rediscuss, in the context of this elliptic interpretation, the
conceptual issues raised in the recent literature. In particular, we
would like to readdress the problem of defining an S-like matrix in a
quantum gravity theory in asymptotic de Sitter space.

Let us briefly review the puzzles that arise in conventional de Sitter
space.  We have already mentioned observer complementarity. Another
issue is that of holography. We would like to have a holographic dual
description of gravity for all of the various asymptotic
geometries. Recently we have learned to describe string theory in
spacetimes that asymptotically approach an anti-de Sitter geometry.
The AdS/CFT correspondence is by now well-established, and in
principle gives a nice holographic description of string theory in
these backgrounds. In Minkowski space too, there are reasons to
believe that a holographic description may exist that involves
holographic screens at past and future null infinity
\cite{Bousso:1999xy,Bousso:1999cb}. But de Sitter space requires yet
another type of holography, because there is no spatial or null
infinity. Various authors have argued that it should be a kind of
timelike holography, for which the holographic screens are spacelike
surfaces in the asymptotic past or future of global de Sitter
space. Strominger, most notably, has proposed a dS/CFT correspondence
\cite{Strominger} similar to AdS/CFT.

A somewhat confusing aspect of holography in global de Sitter space,
however, is that it has two disconnected boundaries. If we think of
the dual CFT as living on these boundaries, then we have to somehow
compute correlation functions of operators some of which may be
inserted on one boundary, while others may act on another boundary.
Not only is it unclear how to compute such correlation functions, it
is also unclear what their physical interpretation is.

A related problem arises in trying to define the analog of an
S-matrix. In quantum field theory, asymptotic incoming and outgoing
states are properly defined only in the asymptotic regions of
spacetime. But for de Sitter space these regions are spacelike, and
there is no single observer who can determine the states both at past
infinity as well as at future infinity.  Consequently, the matrix
elements of S-like matrices in de Sitter space are not measurable
quantities; they are mere meta-observables, rather than observables.
When one considers quantum gravity in asymptotically de Sitter space,
the situation becomes even more serious.  As has been pointed out by
Witten, the only available pairing between in-states and out-states,
CPT, is used to obtain an inner product for the Hilbert space
\cite{Witten}. There then does not seem to be an additional pairing
between in- and out-states that could be used to arrive at an
S-matrix. As the conventional formulation of string theory is based on
the existence of an S-matrix, the lack of an analog of an S-matrix is
worrisome.

Finally, we come to the question of the de Sitter entropy
\cite{Gibbons:mu}. Conventional global de Sitter space makes it hard
to understand the finiteness of the entropy. For, in the far past, the
asymptotic geometry is that of an enormous sphere, which can be
perturbed in very many ways. The vast majority of these perturbations
do not lead to a spacetime that is asymptotically de Sitter in the
future; instead, singularities and black holes form. How the finite
number of states that do lead to asymptotically de Sitter in the
future are characterized is still a mystery.

This paper is organized as follows. In section 2 we briefly describe
de Sitter space and point out, by way of motivation, some facts about
de Sitter space that support the proposed $\ZZ$ identification. In
section 3, we define Schr\"odinger's antipodal identification, and
refine it to include CPT. We then discuss its classical properties and
show that elliptic de Sitter space does not suffer from any obvious
problems, such as closed timelike curves. Next, in section 4, we
consider quantum fields propagating in this space. In particular, we
discuss the vacuum state in the Fock space of a free scalar field. In
section 5, we consider holography. It is here that the advantages of
the elliptic interpretation are perhaps most evident; conceptually,
the holographic theory seems to have a more natural interpretation
with the $\ZZ$ identification than without. In section 6, we discuss
how elliptic de Sitter space might be realized in string theory. We
conclude in section 7.

\section{Mirror Images in de Sitter Space}
\label{sec:dS}

Empty de Sitter space is the unique spacetime with maximal symmetry and
constant positive curvature. In $D$ spacetime dimensions, it is locally
characterized by
\begin{equation}
R_{ab} = {D-1 \over R^2} g_{ab} \; ,
\end{equation}
where $R$ is the radius of curvature of de Sitter space, and by the
vanishing of the Weyl tensor. The cosmological constant $\Lambda$ is
a function of $R$. With the local geometry fixed, the only remaining
freedom lies in choosing the global topology.

It is convenient to think of de Sitter space as a timelike hyperboloid
embedded in \dimplus{D}{1} Minkowski space. The embedding equation is
\begin{equation}
  \label{embed}
  -X_0^2 + X_1^2 + \ldots + X_D^2 = R^2 \; ,
\end{equation}
where $X_I$ are Cartesian coordinates in Minkowski
space. Eq.~(\ref{embed}) makes the $O(1,D)$ isometry group of de
Sitter space manifest. Note that $O(1,D)$, the Lorentz group in
$D\!+\!1$ spacetime dimensions, has four disconnected
components. These are the proper orthochronous Lorentz group and its
composition with the discrete symmetries of P and T, i.e. with parity
and time-reversal. By parity we will always mean a reflection in a
hyperplane of one spatial codimension rather than spatial inversion
through the origin; the discussion is therefore unaffected by whether
the spacetime dimension is odd or even.

For a given point on de Sitter space at embedding coordinate $X$, we
define the {\em antipodal point} to be the point obtained by
reflection through the origin of Minkowski space, i.e. the point with
embedding coordinate $-X$. We then define {\em elliptic de Sitter
space} to be the spacetime in which for every physical event at any
point on de Sitter space there is a CPT-conjugate event at the
antipodal point. Hence we are using our freedom of topology to impose
a $\ZZ$ identification of de Sitter space. Note that
the connected part of the isometry group remains unchanged after
the identification; the $\ZZ$ identification mods out by a center of
the de Sitter group. The preservation of all local isometries
justifies the appellation ``de Sitter space.''

In the remainder of this section, we consider various properties of
global de Sitter space that suggest that information on one side of
the horizon is mirrored on the other side. We do not claim that de
Sitter space {\em must} be antipodally identified; rather, the
examples should be seen as circumstantial evidence that elliptic de
Sitter space may be more natural than global de Sitter space. For a
detailed description of the classical properties of de Sitter space
see \cite{HawkingEllis}; for a recent review see
\cite{Spradlin:2001pw}.

\subsection{Mirror singularities}
\label{sec:mirror-sing}

The great circles, or geodesics, of a sphere are determined by the
intersection of the sphere with planes that pass through the
origin. Similarly, the spatial geodesics of de Sitter space can be
obtained by intersecting it with spacelike planes through the origin of
Minkowski space. It is clear then that {\em every} spatial geodesic
that passes through a point must also pass through its antipode,
because if $X$ lies in a plane through the origin then so does
$-X$. These geodesics form ellipses which are related to each other by
de Sitter transformations. If we think of null rays as degenerate spatial
geodesics, and if we allow them to ``bounce off'' null infinity, then
{\em all} light rays leaving a point converge on the antipodal
point. This last fact affects the singularity structure of Green's
functions of quantum fields.

Consider a scalar field in de Sitter space. It is convenient to
express de Sitter-invariant equations in terms of a dimensionless de
Sitter-invariant variable, $Z$. We can define such a variable by
\begin{equation}
Z(X,Y) = {1 \over R^2} {X \cdot Y}   \; ,
\end{equation}
where the dot product is given by the Minkowski metric. Obviously $Z$
is Lorentz-invariant in $D\!+\!1$ dimensions, and therefore de
Sitter-invariant in $D$ dimensions. For points that are connected by
geodesics, $R \arccos Z$ corresponds to the geodesic distance. In
particular, for any given $X$ if $Y$ is on the light-cone of $X$, then
$Y = X+N$ with $N^2 = 0$. Since $X$ and $Y$ must both lie on the same
de Sitter hypersurface, $X^2 = Y^2 = R^2$, and therefore $Z = +1$. On
the other hand, if $Y$ is on the light-cone of the antipodal point, $Y
= -X + N$ and so here $Z$ takes the value $-1$.

The wave equation for a massive scalar field written in terms of $Z$ is
\begin{equation}
\lf (1 - Z^2) {d^2 \over dZ^2} - DZ{d \over dZ} - m^2/R^2 \rt \phi(Z) = 0 \; .
\label{waveeqn}
\end{equation}
The Wightman functions obey this equation. The precise form of the
solution, a hypergeometric function, is immaterial; the key point is
that it is singular at $Z = 1$. This is analogous to the usual
short-distance singularity at $\sigma = 0$ that one has in Minkowski
space along the light-cones. But now the wave equation is symmetric
under $Z \to -Z$. Therefore in de Sitter space there is a second
solution to Eq.~(\ref{waveeqn}) with a singularity at $Z = -1$,
i.e. on the light-cones of the antipode. Hence we see that, in
contrast to Minkowski space, singularities of Green's functions in de
Sitter space seem to come in pairs. The mirror singularity along the
antipodal light-cones is our first example of duplication in de Sitter
space.

\subsection{Mirror black holes}
\label{sec:mirror-black-holes}

As a second example, consider a Schwarzschild-de Sitter black hole in
$D=d\!+\!1$ spacetime dimensions. The line element has the form
\begin{eqnarray}
\label{eq:1}
ds^2 &=& - F(r) dt^2 + F^{-1}(r) dr^2 + r^2 d\Omega_{d-1}^2 \; , \\
F(r) &=& 1 - {2M \over r^{d-2}} - {r^2 \over R^2} \; .
\end{eqnarray}

If $0< M <M_{\rm max}$\footnote{$M_{\rm max} = \frac{1}{d} \left(
      \frac{(d-2) (d-1)}{2\Lambda} \right) ^{\frac{d-2}{2}}$ is the
      maximal mass. At this value the black hole and cosmological
      horizons coincide.}, there are two horizons: a cosmological
      horizon at $r=r_c$ and a black hole horizon at $r=r_{bh}$, where
      $r_c > r_{bh}$. We will show that, when the solution is
      analytically extended, there is a mirror black hole on the other
      side of the cosmological horizon. Let us introduce
      Kruskal-Szekeres type coordinates and analytically continue the
      metric beyond the cosmological horizon. Note that a priori the
      coordinates in Eq.~(\ref{eq:1}) are only valid for $r_{bh} < r <
      r_c$.

In terms of its roots, the function $F(r)$ can be written as
\begin{equation}
F(r) = - {1 \over R^2 r^{d-2}} (r-r_c) (r-r_{bh}) \prod_{n=3}^{d}
(r-r_n) \; ,
\end{equation}
where $r_c$ and $r_{bh}$ are the only real positive roots. Hence
\begin{equation}
F^{-1}(r) = \frac{c_1}{r-r_c} + \frac{c_2}{r-r_{bh}} + \sum_{n=3}^{d}
\frac{c_n}{r-r_n} \; ,   \label{eq:3}
\end{equation}
for certain constants $c_n$. Define Eddington-Finkelstein coordinates through
\begin{equation}
  dx^{\pm} = dt \pm \frac{dr}{F(r)} \; ,
\end{equation}
which, using Eq.~(\ref{eq:3}), is easily integrated to give
\begin{equation}
  \label{eq:5}
  x^{\pm} = t \pm \left\{ c_1 \ln{(r-r_c)} + c_2 \ln{(r-r_{bh})} +
  \sum_{n=3}^{d} c_n \ln{(r-r_n)} \right\} \; .
\end{equation}
In terms of these coordinates, the metric takes the form
\begin{equation}
  ds^2 = -F(r) dx^+ dx^- + r^2 d\Omega_{d-1}^2 \; .
\end{equation}
Finally, introduce Kruskal-Szekeres coordinates through
\begin{equation}
\begin{split}
  \label{eq:7}
  U &= e^{-\frac{x^-}{2c_1}} \\
  V &= -e^{\frac{x^+}{2c_1}} \; ,
\end{split}
\end{equation}
where it is clear that $U>0$ and $V<0$. The metric becomes
\begin{equation}
ds^2 = 4c_1^2 \, \frac{F(r)}{UV} dU dV + r^2(U,V) d\Omega_{d-1}^2 \; .
\end{equation}
In terms of these coordinates the metric is regular at $r=r_c$ and we
can analytically continue to the full range $-\infty < U,V <
\infty$. Note from Eqs.~(\ref{eq:5}) and~(\ref{eq:7}) that $r(U,V) =
r(UV)$ and thus $F(r)=F(UV)$. Hence, if $F(UV)$ is zero for certain
nonzero values of $U$ and $V$, e.g. at the black hole horizon, then it
will also be zero at $-U$ and $-V$. This second horizon is antipodal
from the first and thus we find that black holes in de Sitter space
come in antipodal pairs. Actually this is a choice: instead of
extending the metric analytically entirely to the other side, we could
have replaced the antipodal black hole by a static, spherically
symmetric mass distribution with the same total mass.

Now consider adding charge to the de Sitter black hole
\cite{Romans}. De Sitter space cannot support Noether charges because
its spatial sections are compact. The total charge has to add up to
zero; the antipodal black hole therefore necessarily carries equal,
but opposite charge. Moreover, for the same reason there cannot be any
net angular momentum. This leads us to propose that the antipodal map
must be combined with charge conjugation, C.

\section{The Elliptic Interpretation of de Sitter Space}
\label{sec:elliptic}

The elliptic interpretation of de Sitter space consists of identifying
points that are related by the antipodal map
\begin{equation}
X^I\to -X^I \; ,
\end{equation}
with $I=0,1,\ldots, D$, together with charge conjugation, C. We will
see that this means that particles and/or events at $X^I$ and $-X^I$
are related by CPT. We thus have an involution, a $\ZZ$ map. The fixed
point of the map, $X^I = 0$, is not itself in de Sitter space, so this
is a freely-acting symmetry. The quotient space, $dS/\ZZ$, is
therefore a homogeneous space with no special points.

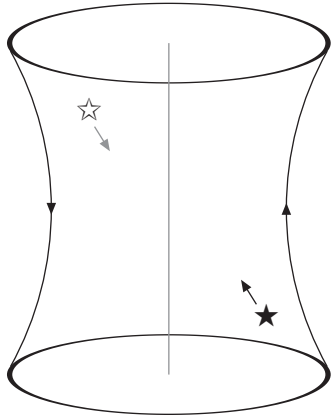
\begin{figure}[hbtp]
        \centering
\begin{picture}(0,140)
\Oval(0,132.5)(15,60)(0)
\Oval(0,7.5)(15,60)(0)
\CArc(-167,70)(122.71,-29,29)
\CArc(167,70)(122.71,151,209)
\ArrowLine(-44.29,70.01)(-44.29,69.99)
\ArrowLine(44.29,69.99)(44.29,70.01)
\put(32,26){\ding{72}}
\LongArrow(33,34)(28,42)
\SetColor{Gray}
\definecolor{dotgray}{gray}{0.4}
\Line(0,7.5)(0,132.5)
\put(-35,104){\ding{73}} 
\LongArrow(-28,101)(-23,93)
\SetColor{Black}
\end{picture}
\caption[]{\small The antipodal map reverses the local arrow of time.}
      \label{fig:antipode}
\end{figure}
Note that the antipodal map also inverts the direction of time; see Figure
\ref{fig:antipode}. For example, consider global coordinates. The line element
reads
\begin{equation}
ds^2=-dT^2+R^2\cosh^2 (T/ R) \lf d\theta^2 + \sin^2\theta~
d\Omega^2_{D-2} \rt \; .
\end{equation}
In these coordinates the antipodal map is given by
\begin{equation}
T\to-T\qquad \theta\to\pi-\theta\qquad \Omega\to \Omega^A \; ,
\end{equation}
where $\Omega^A$ are the angular coordinates of the point antipodal on the
\dimmin{D}{2} sphere to the point labeled by $\Omega$, and time is reversed,
$T \to -T$. In the rest of this section, we show that elliptic de Sitter space
is nevertheless classically consistent, with no problems of causality or
closed timelike curves. We will also demonstrate that the map between a
particle and its antipodal image is CPT.

\subsection{Causality}
\label{sec:causality}

The antipodal map identifies points at positive $T$ with points at
negative $T$, and so one may wonder whether there are probems with
causality or closed timelike curves. That such problems do not arise
was explained by Schr\"odinger \cite{Schroedinger}. We just give here
our version of the argument.

First, let us go to the embedding space. It is easily seen that two
antipodal points at $X$ and $-X$ are always spacelike separated, since
$X^2=R^2>0$. Moreover, the intersection of the two light-cones
that start at antipodal points never intersect the de Sitter
hypersurfaces, because if $Y$ is the embedding coordinate of a common
point on the light-cones emanating from $X$ and $-X$, then
\begin{equation}
(Y+X)^2=(Y-X)^2=0 \To Y^2=-R^2 \; ,
\end{equation}
so $Y$ does not lie on the de Sitter hypersurface. This means that the
light-cones of two antipodal points within de Sitter space do not
intersect. Therefore a pair of events that take place at antipodal
points cannot both influence the same event in their past and
future. In particular, there are no closed timelike curves after $\ZZ$
identification.

What about closed null curves? A point on $\scriminus$ is connected by
a lightlike trajectory to its antipodal image on $\scriplus$. So at
first this appears to give rise to an infinity of closed lightlike
trajectories. However, these light-rays do not constitute closed
trajectories in de Sitter space for three important reasons. First of
all, ``points'' at $\scriplus$ and $\scriminus$ are not really points
in de Sitter space. They have to be added as points at ``infinity,''
and so they are only part of a formal compactification of de Sitter
space. De Sitter space itself is noncompact and does not include these
points. A second, related reason is that the affine parameter along
the seemingly closed lightlike trajectory is actually infinite,
essentially because the points are at $\scri$. Finally, a third reason
that the lightlike trajectory is not really closed, is that one cannot
continue along the trajectory a second time, third time, etc. without
reversing direction each time one is at the endpoints on $\scriplus$
or $\scriminus$. This is not what happens on a usual closed
trajectory, such as on a timelike $S^1$.

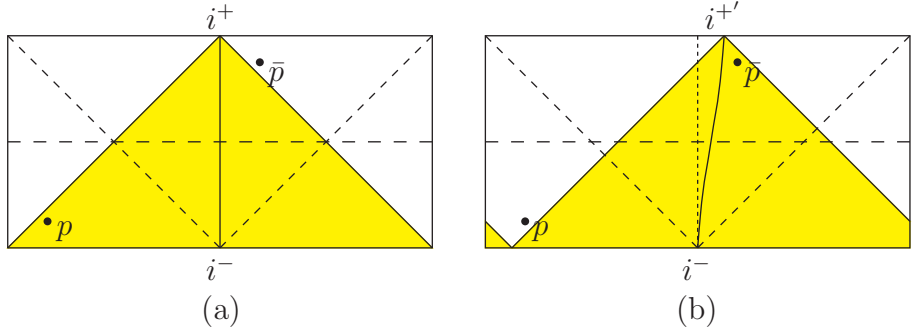
\begin{figure}[hbtp]
\centering
\begin{picture}(0,110)
\SetScale{1}
\CTri(-90,100)(-170,20)(-10,20){Yellow}{Yellow}
\CTri(20,20)(170,30)(170,20){Yellow}{Yellow}
\CTri(20,20)(100,100)(170,30){Yellow}{Yellow}
\CTri(10,30)(20,20)(10,20){Yellow}{Yellow}
\EBox(-170,20)(-10,100)
\EBox(10,20)(170,100)
\DashLine(90,20)(90,100){1.5}
\Text(-90,2)[t]{(a)}\Text(90,2)[t]{(b)}
\Text(-90,17)[t]{$i^-$}
\Text(90,17)[t]{$i^{-}$}
\Text(-90,103)[b]{$i^+$}
\Text(100,103)[b]{$i^{+'}$}
\Line(-90,20)(-90,100)
\Curve{(90,20)(91,32)(95,60)(99,88)(100,100)}
\Line(-90,100)(-170,20)
\Line(-90,100)(-10,20)
\Line(100,100)(20,20)
\Line(100,100)(170,30)
\Line(10,30)(20,20)
\DashLine(-90,20)(-10,100){3}
\DashLine(-90,20)(-170,100){3}
\DashLine(90,20)(170,100){3}
\DashLine(90,20)(10,100){3}
\DashLine(-170,60)(-10,60){5}
\DashLine(10,60)(170,60){5}
\Vertex(-155,30){1.5} \Vertex(-75,90){1.5}
\Text(-152,30)[lt]{$p$} \Text(-72,90)[lt]{$\bar{p}$}
\Vertex(25,30){1.5} \Vertex(105,90){1.5}
\Text(28,30)[lt]{$p$} \Text(108,90)[lt]{$\bar{p}$}
\end{picture}
\caption[]{\small These Penrose diagrams of de Sitter space have been
opened up to make all antipodal points distinct. The left and right
edges of a diagram are identified, and every point in the interior
(except on the central vertical line) now signifies an
$\mathbb{R}P^{D-2}$, instead of a $S^{D-2}$. The antipode of a given
point is reached by reflecting about the dashed horizontal line, and
moving horizontally by half the width of the diagram. Two antipodes,
marked $p$ and $\bar{p}$, are shown. In (a) an observer traveling from
$i^-$ to $i^+$ has $p$ but not $\bar{p}$ in his causal past (shaded),
while in (b) an observer with a different worldline can see $\bar{p}$
but not $p$. The antipodal image of a shaded region is the unshaded
region, giving every observer complete information after the $\ZZ$
identification.}
\label{fig:obs-desitter}
\end{figure}

It is also useful to analyse the antipodal identification from the
point of view of inertial observers.  All points inside the causal
diamond of an observer have antipodal points outside the causal
diamond. The antipodal points belong to the causal diamond of the
antipodal observer, on the inaccessible ``dark side of the moon.''
Therefore exactly one of every pair of antipodal events is
observable. Which event of each pair is observed depends on the
location of the observer; see Figure \ref{fig:obs-desitter}. For
example, the observer living at the south pole will see precisely all
antipodal images of the events that his colleague at the north pole
sees. Other observers will see something in between, namely for some
part ``northern'' events, and for the rest ``southern'' events, but
every event is observed once and no more than once.

What about events that take place outside the causal diamonds of the
observer at the south and the north pole? These are the events that
take place at the upper and lower parts of the Penrose diagram near
past and future infinity. In the elliptic interpretation of de Sitter
space these upper and lower regions are identified. The usual square
Penrose diagram for de Sitter space is somewhat misleading in the
sense that it seems to indicate that all points in the upper region
are in the causal future of points of the lower region. But one has to
remember that every point represents a \dimmin{D}{2} sphere, and
points that are identified by the antipodal map are on opposite sides
of these spheres. A clearer way to see the causal structure of
elliptic de Sitter space is to represent the \dimmin{D}{2} spheres as
two points, each of which is a real projective sphere; see Figure
\ref{fig:obs-desitter}. Now one can see that a geodesic that connects
two identified points in the upper and lower regions has to travel
forward in time, but also has to go around the sphere.  Since all
antipodal points are spacelike separated, the resulting geodesic is
indeed spacelike.

Next consider the horizon itself. Without loss of generality we may
consider an observer at the ``north pole'' $\theta=0$ of the spatial
\dimmin{D}{1} sphere $S^{D-1}$. His past and future event horizon are
given by $\theta = 2 \arg(i+e^{\pm {T\over R}})$, and intersect at $T
= 0$ at the equator of his \dimmin{D}{1} sphere, described by the
\dimmin{D}{2} sphere at $\theta= \pi/2$. The intersection takes place
at the midpoint of the square Penrose diagram. Therefore only by
sending a signal at $T=-\infty$ can he contact the equator in time for
a signal to come back to him precisely at $T=\infty$.  Hence, if we
exclude the points at infinity, there is no way that the observer can
communicate (sending a question and getting a reply) with points on
the equator. Events that happen right on the equator are identified
with the events that happen at the antipode of the equator itself. But
this fact only becomes apparent to the observer at the north pole (or
south pole) at $T=\infty$ (or $T=-\infty$). We conclude that at no
finite time can any observer ever directly detect the duplication of
events in elliptic de Sitter space.

Finally, note that the asymptotic geometry of elliptic de Sitter space
consists of a single $S^{D-1}$, since the $\ZZ$ identification maps
$\scriplus$ and $\scriminus$ to each other. This property will be
useful when we consider the holographic theory.

\subsection{CPT}
\label{sec:CPT}

Any two antipodal points can be mapped to the north and south pole
corresponding to $X^D=\pm R$, $X^k=0$ for $k=0,1,\ldots, D\!-\!1$.
Without loss of generality, consider a particle with trajectory
$X^I(\tau)$, $I=0,1,\ldots,D$ in the embedding space passing through
the north pole at $\tau=0$. Its antipodal image is $-X^I(\tau)$ and
passes through the south pole. Let us apply time-reversal to the
antipodal image:
\begin{equation}
T: -X^I(\tau) \to -X^I(-\tau) \; .
\end{equation}
The relativistic momentum of the particle at the north pole is
$p^I=\dot{X}^I$. Note that $p^D=0$ at $\tau = 0$. At the south pole
the momentum is also given by $p^I$ since it is $-X^I(-\tau)$
differentiated with respect to $\tau$ at $\tau=0$. So in the embedding
space the momentum is pointing in the same direction. However, in
order to compare this to the momentum at the north pole, we have to
parallel transport the vector from the south pole to the north
pole. There are many ways of doing this because there are an infinite
number of spatial geodesics passing through both the north and the
south pole. Let us pick one of them, say the one that appears when we
intersect de Sitter space with the two-dimensional plane $X^m=0$ for
$m=0,1,\ldots,D\!-\!2$. This gives as a geodesic $X^{D-1}=R\sin\theta$,
$X^D=R\cos\theta$. At $\theta=0$ we are at the north pole, at
$\theta=\pi$ at the south pole. Parallel transport of the momentum
$p^I$ along this trajectory gives a momentum $(p^\prime)^I$ which
satisfies
\begin{equation}
\begin{split}
(p^\prime)^m & =p^m \; , \qquad m=0,\ldots, D\!-\!2 \\
(p^\prime)^{D-1} & =-p^{D-1} \; .
\end{split}
\end{equation}
We see that one of the spatial components of the momentum has changed
sign. That is the result of a reflection in a $D\!-\!2$ spatial
dimensional hyperplane. Thus it corresponds to parity, even though in
the embedding space $X^I \to -X^I$ corresponds to an inversion. The
plane of reflection in this case is the plane $X^{D-1}=0$. Had we
chosen a different geodesic it would have been another plane. Note
that this is consistent, because parallel transport along two
different geodesics differs by a rotation, equal to the integrated
curvature between the geodesics. This is precisely what one finds if
one composes the two reflections in the planes associated with those
geodesics.

Therefore going around from a point in de Sitter space to its
antipodal point has the effect of acting on the tangent space by
PT. Since our $\ZZ$ map also requires that we act with
charge-conjugation, C, the cumulative effect is to relate antipodal
points by CPT.

\subsection{The arrow of time}
\label{sec:arrow-of-time}

The antipodal map, $X^I\to -X^I$, changes the sign of the time
coordinate of the embedding space, and also that of the direction of
time in de Sitter space. The resulting quotient space, $dS/\ZZ$, is as
a consequence not time-orientable: although one can locally
distinguish past and future, there is no global direction of
time. This fact clearly changes many standard notions about space and
time that we are accustomed to. For instance, it is impossible to
choose a Cauchy surface for elliptic de Sitter space that divides
spacetime into a future and a past region.

Since the microscopic laws of physics are generally time-reversible,
that is CPT-invariant, there is no problem with time unorientability
at a microscopic level. It is more subtle, however, to formulate
macroscopic laws of physics on a time unorientable spacetime. For
example, the evolution of stars clearly shows a direction of time; one
never observes a neutron star turning into a massive star through the
enormous implosion of a stellar envelope, yet this is what the
antipodal image of a type II supernova would look like.

For sufficiently simple situations, a single observer can always
choose a preferred direction of time in the observable part of the
universe, consistent with the second law of thermodynamics. Consider
an isolated thermodynamic system in configuration $A$, with antipodal
image $A'$, which evolves into system $B$, with antipodal image
$B'$. If the entropies are such that $S(B) \gg S(A)$, an observer who
observed both $A$ and $B$ would say that $A$ preceeded $B$. Since the
primed and unprimed systems have the same entropy, this would mean
that an observer who observed both $A'$ and $B'$ would say that $A'$
preceeded $B'$, and would therefore have time flowing in the opposite
direction. Finally an observer who saw, say, $A$ and $B'$ would see
them as two distant, spacelike-separated systems, rather than one
system evolving into another. For this observer the choice of arrow of
time is independent of the relative entropies of the two systems. In
this simple scenario, no problems arise for any observer.

However, now consider a second thermodynamic system in states $C$ and
$D$. For example, $A, B$ and $C, D$ could describe the configurations
before and after two supernova explosions. It is easy to check that if
both $C$ and $D$ are outside the past light-cone of $B$, then there is
always at least one observer who witnesses a dramatic violation of the
second law, irrespective of his choice of time arrow. This is not
fatal because, after all, the underlying dynamics do enjoy a CPT
symmetry. Rather, the issue is of what the allowable initial
conditions are. One consistent treatment is to say that there are
simply no highly-ordered systems present. (This would include,
unfortunately, realistic observers...) Indeed, there are reasons to
believe that our observed macroscopic arrow of time may be related to
boundary conditions at cosmological singularities. It would be very
interesting to see if there are cosmological scenarios \cite{aguirre}
that can be built out of elliptic de Sitter space.

An alternate and quite different viewpoint is to argue that before one
can even assign events in spacetime, one should first choose an
observer. Indeed, even classically, different observers can have
rather different interpretations of local physics, as happens in the
membrane paradigm for black holes \cite{paradigm,stu,memaction}. Then
for a given observer one can always arrange events to be consistent
with his preferred arrow of time. One only runs into trouble if one
tries to consider many observers, who all choose a preferred time
direction. But such considerations are against the notion of observer
complementarity, which forbids simultaneous consideration of observers
on opposite sides of an event horizon.

\subsection{The $\Lambda \to 0$ limit}
\label{sec:lambda-to-0-limit}

An interesting limit of de Sitter space is the limit in which the
cosmological constant is sent to zero, so that spacetime locally
becomes Minkowski space. This limit has to be treated with care; the
quantities of interest should vary smoothly as $\Lambda \to 0$. For
elliptic de Sitter space, the $\Lambda \to 0$ limit seems
sensible. The causal properties of the $\ZZ$ quotient space for any
finite $\Lambda$ are similar to that of Minkowski space, in the sense
that every observer who waits long enough has the chance to observe
(and emit signals to) any event in spacetime, just as in Minkowski
space. The main difference is that elliptic de Sitter space is not
time-orientable. However, as the cosmological constant goes to zero
this difference disappears to the null boundaries.

Now, if elliptic de Sitter space goes to Minkowski space in this
limit, it seems to imply that global de Sitter, being its two-fold
cover, in fact goes to {\em two} copies of Minkowski space, where the
second copy is the CPT-conjugate of the first. The significance of
these remarks will be more clear once we discuss the de Sitter analog
of an S-matrix which, we will argue, exists in elliptic de Sitter
space but does not appear to exist in global de Sitter space.

\section{Quantum Fields in Elliptic de Sitter Space}
\label{sec:bulk-fields}

In this section, we study the quantization properties of a scalar
field propagating in elliptic de Sitter space. Some aspects of the
quantum field theory of a free scalar field in elliptic de Sitter
space have previously been discussed in
\cite{Gibbons,Sanchez}\footnote{After this work had been posted, a
paper \cite{Banks} related to this section appeared on the archive.}.

Elliptic de Sitter space is not simply connected; there are closed
spacelike curves going from a point to the antipodal point that are
noncontractible. Therefore tensor fields on elliptic de Sitter space
can be sections of a twisted bundle over spacetime. Since the first
homotopy group is $\pi_1(dS_n/\ZZ) = \ZZ$, we can essentially choose a
sign for the phase of a tensor field as the field is carried around a
noncontractible loop. Consider then a complex scalar field. We can
choose either periodic or antiperiodic boundary conditions. If we
choose periodic conditions, the condition a complex field must satisfy
takes the form
\begin{equation}
\label{eq:per-cond}
\Phi_{\pm}(\bar{x}) = \pm \Phi_{\pm}^*(x) {\; ,}
\end{equation}
where $\bar{x}$ denotes the antipodal point to $x$, and the subscript
$\pm$ indicates whether we have chosen periodic or anti-periodic
boundary conditions. If we write $\Phi_{\pm}(x) = \Phi_1(x) + i
\Phi_2(x)$, then the real and imaginary parts have periodic
(antiperiodic) and antiperiodic (periodic) boundary conditions
respectively for the plus (minus) subscript.

Globally, one can expand a scalar field in terms of ``Euclidean''
modes. These are field configurations that satisfy the wave equation,
with boundary conditions that are such that the modes can be
analytically continued from the spherical harmonics on a sphere. A
property of the Euclidean modes is that they can be chosen to obey
\begin{equation}
\label{euclidean}
\phi_n^E(\bar{x}) = {\phi_n^E}^*(x) {\; ,}
\end{equation}
and we will assume that our modes satisfy this condition. Normally, one
expands the field in terms of its modes as
\begin{equation}
\Phi_{\pm}(x) = \sum_n [ a_{n,\pm} \phi_n^E(x) + a_{n,\pm}^\dagger
{\phi_n^E}^*(x) ] {\; .}
\end{equation}
In elliptic de Sitter space, however, the field must additionally obey
the periodicity condition Eq.~(\ref{eq:per-cond}). This implies that
\begin{equation}
\label{eq:11}
a_{n,\pm}^\dagger = \pm a_{n,\pm} {\; ,}
\end{equation}
indicating that the global quantization scheme breaks down. As a
result, a global Fock space no longer exists; any creation operator
acting on a vacuum state would annihilate it. Intuitively, this
happens because the identified spacetime is not time-orientable.
Creation and annihilation operators create and destroy quanta of
positive energy, but if the spacetime is not time-orientable positive
energy cannot be defined globally. For essentially the same reason,
the inner product of modes over a spatial slice, $\Sigma$, through
elliptic de Sitter space always gives zero. This is because the
Klein-Gordon inner product
\begin{equation}
(\phi_m, \phi_n) = -i\int_\Sigma (\phi_m \pl_t \phi^*_n -\phi^*_n \pl_t \phi_m)
\end{equation}
vanishes as a consequence of the flipping of the direction of time.

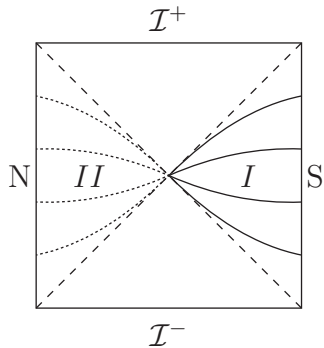
\begin{figure}[hbtp]
\centering
\begin{picture}(0,140)
\EBox(-50,20)(50,120)
\DashLine(-50,20)(50,120){3}
\DashLine(-50,120)(50,20){3}
\Text(0,125)[b]{$\mathcal{I}^+$}
\Text(0,15)[t]{$\mathcal{I}^-$}
\Text(-52,70)[r]{N}
\Text(52,70)[l]{S}
\Curve{(0,70)(25,50)(50,40)}
\Curve{(0,70)(25,62)(50,60)}
\Curve{(0,70)(25,78)(50,80)}
\Curve{(0,70)(25,90)(50,100)}
\DashCurve{(-50,40)(-25,50)(0,70)}{1}
\DashCurve{(-50,60)(-25,62)(0,70)}{1}
\DashCurve{(-50,80)(-25,78)(0,70)}{1}
\DashCurve{(-50,100)(-25,90)(0,70)}{1}
\Text(30,70)[]{$I$}
\Text(-30,70)[]{$II$}
\end{picture}
        \caption[]{\small Penrose diagram of de Sitter space. Region $I~(II)$
          corresponds to the static patch of an observer on the south (north)
          pole. The solid lines indicate equal time slices in the static time,
          they are Cauchy surfaces for region $I$. The dotted lines are their
          antipodal images, and constitute Cauchy surfaces for region $II$.
          When a solid line is continued through the horizon, onto its
          antipodal image, it constitutes a Cauchy surface for the whole
          space.}
      \label{fig:cauchy}
\end{figure}

The vanishing of the norm and the lack of a nontrivial Fock space may
seem like serious afflictions, but actually in elliptic de Sitter
space it is more natural to build a Fock space with oscillators
defined on a static patch. To see this, note that under the antipodal
identification Cauchy surfaces for the static patch constitute Cauchy
surfaces for the whole space, as shown in Figure
\ref{fig:cauchy}. Consider the static patch associated with an
observer at the south pole, region $I$ in Figure \ref{fig:cauchy}. In
this region there is a well-defined direction of time (except
precisely at the horizon) and Fock space operators,
$a_\omega^{(\dagger)I}$, can consequently be defined.
The vacuum is then defined in the usual way,
\begin{equation}
a_\omega^I |\mbox{vac} \rangle = 0 \; , \qquad \forall \omega>0 {\; ,}
\end{equation}
and a Fock space can be constructed. The antipodal map identifies
\begin{equation}
a_\omega^{I(II)} \leftrightarrow a_\omega^{\dagger II(I)} {\; ,}
\end{equation}
i.e. creation (annihilation) operators in region $I$ are identified
with annihilation (creation) operators in region $II$;
cf. Eq.~(\ref{eq:11}). It would be interesting to work out the behavior
of higher-spin fields and, in particular, fermions in elliptic de
Sitter space.

Different observers are related by Bogolubov transformations. These
are invertible, mapping pure states onto pure states. We expect no de
Sitter-invariant pure states; in particular the vacuum state is not
invariant, as is obvious by considering observers that are antipodal
to each other. There are nevertheless de Sitter-invariant mixed
states. These states correspond to de Sitter-invariant pure states in
the global Fock space, traced over the modes behind the horizon. In
particular, there is a state that is observed as a thermal state by
any observer moving along a timelike geodesic $x(\tau)$. To see this,
consider a real scalar field on the identified spacetime, given in
terms of a scalar field on the unidentified space as
\begin{equation}
\Phi_\pm(x) = \frac{1}{\sqrt{2}} ( \Phi(x) \pm \Phi(\bar{x}) ) {\; .}
\end{equation}
This field satisfies the condition Eq.~(\ref{eq:per-cond}) for a real
field. The Wightman function takes the form \cite{Sanchez}
\begin{equation}
\label{eq:13}
G^0_\pm (x(\tau),x(\tau')) = G^0(x(\tau),x(\tau')) \pm
G^0(x(\tau),\overline{x(\tau')}) {\; ,}
\end{equation}
where $G^0(x,x')$ is the Euclidean Green's function on the
unidentified de Sitter space. In obtaining this we have used the fact
that $G(x,x') = G(\bar{x},\bar{x}')$, which holds because under $x \to
\bar{x}$ and $x' \to \bar{x}'$, the de Sitter-invariant quantity $Z$
remains unchanged, and $G^0(x,x')$ is a function only of $Z(x,x')$
(see section \ref{sec:mirror-sing}) since the Wightman functions are
de Sitter-invariant. Assuming, without loss of generality, that the
observer remains static on the south pole, $Z(x(\tau),x(\tau'))$ is
given in terms of static coordinates by $\cosh((\tau-\tau')/R)$ when
$\tau$ is the proper time. The Green's function thus takes the form
\begin{equation}
G_\pm(x(\tau),x(\tau')) = G^0(\cosh((\tau-\tau')/R)) \pm
G^0(-\cosh((\tau-\tau')/R)) {\; .}
\end{equation}
This is a thermal Green's function at a temperature $1/2\pi R$. So
even though every observer in elliptic de Sitter space has complete
information, one still has thermal states at the de Sitter
temperature. This is because thermal emission of particles (which can
be viewed as quanta that have tunneled through the horizon) is a
process which only requires half of global de Sitter space
\cite{newcoords,medved,tunnel}. Unlike the unidentified case, however,
there is no frame for which this Green's function corresponds to a
pure vacuum state.

As discussed in \cite{Chernikov,Tagirov,Bousso:2001mw,Spradlin}, there
is a one-parameter family of de Sitter-invariant Green's functions in
unidentified de Sitter space, parametrized by $\alpha$, with the
Euclidean Green function corresponding (in the parametrization of
\cite{Allen}) to $\alpha = 0$. The existence of such a family stems
from the fact that on de Sitter space one can add an antipodal source,
as we saw in section \ref{sec:mirror-sing}. The corresponding modes
are related by Bogolubov transformations:
\begin{equation}
\phi_n^\alpha (x) = \cosh \alpha \phi^E_n (x) + \sinh \alpha
\phi^E_n(\bar{x}) \; . 
\end{equation}
By Eq.~\ref{euclidean}, the new modes mix the old positive and
negative energy modes and therefore define a new, inequivalent
vacuum. The $\alpha$-vacua, $| \alpha \ket$, called Mottola-Allen
states \cite{Mottola,Allen}, form a one-parameter family of de
Sitter-invariant vacua. Presumably, they correspond to (nonthermal) de
Sitter-invariant states on the elliptically identified space. The
$\alpha$-vacua have Green's functions given by
\begin{equation}
G^\alpha (x,x') = \bra \alpha | \Phi(x) \Phi(x') | \alpha \ket \; .
\end{equation}
Substituting the mode expansion and the Bogolubov transformation for a
field satisfying Eq.~\ref{eq:per-cond}, the $\alpha$-Wightman function
on the identified space takes the form \cite{Sanchez}
\begin{equation}
G_\pm^\alpha(x,x') = e^{\pm2\alpha} G_\pm^0(x,x') \; ,
\end{equation}
where $G_\pm^0(x,x')$ is given by Eq.~(\ref{eq:13}), which corresponds
to $\alpha=0$. In elliptic de Sitter space the Green's functions for
the different $\alpha$-vacua differ by an overall normalization
(ignoring subtleties involving $i \epsilon$ prescriptions). We regard
the Mottola-Allen states for $\alpha \neq 0$ as unphysical, since
their Green's functions do not have the short-distance singularities
that we expect from Minkowski space. The Green's function on elliptic
de Sitter space has singularities on the light-cone as well as on the
light-cone of the antipode, even for $\alpha = 0$. The singularities
have equal strength but can have a relative plus or minus sign due to
the double-valuedness of the phase.

\section{Holography in Elliptic de Sitter Space}
\label{sec:holography}

Now we turn to the theory on the boundary. An immediate consequence of
taking a $\ZZ$ quotient is that every observer now has access to all
of elliptic de Sitter space. Moreover, the antipodal identification
implies that the spacetime now has only a single spacelike
boundary. Hence the holographic dual theory is a Euclidean conformal
field theory on a {\em single} sphere. In the spirit of the dS/CFT
correspondence we shall consider at first the general features of the
holographic CFT, independent of the details of the theory. The
discussion does not need the corresponding bulk fields to be free;
indeed, it applies also to gravity. We will find that the holographic
properties of elliptic de Sitter space are very good, with satisfying
implications for observer complementarity, the existence of an
S-matrix, and a possible expanation of the finiteness of the de Sitter
entropy.

\subsection{Holographic time evolution}
\label{sec:timeevolution}

Even though we do not know what the interior of quantum de Sitter
space looks like, we can still say the following. Classically, the
past and future light-cones of an observer intersect the \dimmin{D}{1}
spheres at asymptotic infinity on \dimmin{D}{2} spheres. In fact,
after identification both light-cones intersect the {\em same}
sphere. The polar angle at which the light-cones emanating from time
$T$ (at the north pole) intersect the $S^{D-2}$ at $\scri$ is given by
\begin{equation}
\theta(T) = 2 \arctan \lf \tanh {T \over 2} \rt + {\pi \over 2} \; .
\end{equation}
At $T=-\infty$ this is zero. At $T=0$, $\theta = \pi /2$, and at
$T=\infty$ it is $\pi$. So by choosing an $S^{D-2}$ at a certain
radius on $\scri$ we are basically taking the point of view of an
observer who is in the middle of de Sitter space at a certain time
$T$. This is holography at work: we do not need to go to the interior
of de Sitter space to describe time evolution, we do it at the
boundary. Even in the quantum theory, since the metric near the
boundary still looks like classical de Sitter space, and we have the
$SO(1,D)$ de Sitter group acting, we can use the global time $T$ to
measure the distance from the scris to the poles.

Now, time translations increase the distance with respect to the north
pole, and decrease the distance to the south pole. In fact, this is
precisely what scale transformations do. To see this, map the north
pole patch to flat Euclidean space, and similarly for a neighborhood
of the south pole. Then the transition function that glues the two
together is inversion $\vec{x}\to \vec{y}= {\vec{x}\over
|\vec{x}|^2}$, which is a conformal transformation. But now note that
scaling up in $x$ is equivalent to scaling down in $y$, exactly like
time translations in the bulk.

That time evolution in the bulk leads to scale transformations in the
boundary was already emphasized by Strominger \cite{Strominger}. In
planar coordinates covering say the causal past, the line element is
$ds^2 = -dt^2 + \exp(-2t/R) dx^2$, and it follows that
\begin{equation}
t \to t + \lambda \qquad x \to e^{\lm/R} x      \label{isometry}
\end{equation}
is an isometry of the metric. Alternatively, one can use static
coordinates in the upper or lower region of the Penrose diagram. The
line element is
\begin{equation}
ds^2 = (r^2/R^2 -1) dt_s^2 - {dr^2 \over r^2/R^2 - 1} + r^2
d \Omega^2_{D-2} \; ,
\end{equation}
and the ``Hamiltonian'' $\pl /\pl t_s$ is manifestly a Killing
vector. In fact, it generates the same isometry as
Eq.~(\ref{isometry}), as can be seen by transforming to $r = |\vec{x}|
\exp(-t/R)$ and $t_s = t + {1 \over 2} R \ln (r^2/R^2 - 1)$). From the
metric it is clear that this is now a spacelike vector, as indeed it
should be since it now corresponds to dilations of the boundary
sphere. We note in passing that there is, however, an important
difference between the patches covered by these coordinates and
elliptic de Sitter space: the boundary of the inflationary patch has
the topology $\mathbb{R}^{D-1}$, while elliptic de Sitter space has an
$S^{D-1}$, which contains an extra point.

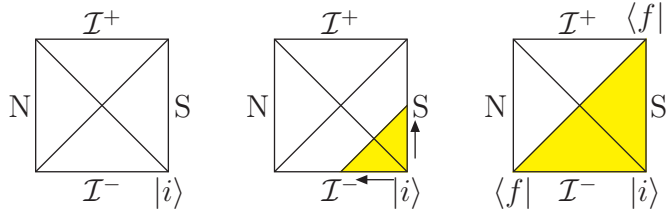
\begin{figure}
 \centering
\begin{picture}(0,70)
\CTri(0,10)(25,10)(25,35){Yellow}{Yellow}
\CTri(65,10)(115,10)(115,60){Yellow}{Yellow}
\Boxc(-90,35)(50,50)
\Boxc(0,35)(50,50)
\Boxc(90,35)(50,50)
\Line(-115,10)(-65,60)
\Line(-65,10)(-115,60)
\Line(-25,10)(25,60)
\Line(25,10)(-25,60)
\Line(0,10)(25,35)
\Line(65,10)(115,60)
\Line(115,10)(65,60)
\Text(-90,62)[b]{$\mathcal{I}^+$}
\Text(-90,8)[t]{$\mathcal{I}^-$}
\Text(-117,35)[r]{N}
\Text(-63,35)[l]{S}
\Text(-65,8)[t]{$|i\rangle$}
\Text(0,62)[b]{$\mathcal{I}^+$}
\Text(0,8)[t]{$\mathcal{I}^-$}
\Text(-27,35)[r]{N}
\Text(27,35)[l]{S}
\Text(25,8)[t]{$|i\rangle$}
\LongArrow(28,15)(28,28)
\LongArrow(20,7)(7,7)
\Text(90,62)[b]{$\mathcal{I}^+$}
\Text(90,8)[t]{$\mathcal{I}^-$}
\Text(63,35)[r]{N}
\Text(117,35)[l]{S}
\Text(115,8)[t]{$|i\rangle$}
\Text(115,62)[b]{$\langle f|$}
\Text(65,8)[t]{$\langle f|$}
\end{picture}
\caption{\small In the far past, an observer at the south pole might describe
the state of the world by an initial state $|i \ket$ on
\scriminus. This evolves in time until it becomes a final state $\bra
f|$ on \scriplus. The antipodal map relates this again to a state on
\scriminus. In- and out-states are therefore associated with a single
surface, as in a conventional CFT.}
\label{fig:time-evolution}
\end{figure}

This leads to a nice picture of how an observer would view the
CFT. Consider an observer in elliptic de Sitter space. By means of de
Sitter transformations, the worldline of any inertial observer can be
mapped to the time axis, say at the south pole. In the far past, such
an observer would characterize the world by an in-state, $|i\ket$. As
in conventional CFT with radial quantization, we would like to assign
incoming states to the origin. Here we choose the origin as the point
where the observer's worldline intersects
$\scriminus$. Correspondingly, we associate an in-state at the south
pole of the boundary sphere. As time passes, the observer moves
vertically along the Penrose diagram. As we have seen this corresponds
to a dilation on the sphere. Finally, in the far future, the observer
describes the world by an out-state, $\bra f|$. This is where the
elliptic interpretation comes in: the out-state is mapped to the
antipodal point on the same $S^{D-1}$ as the in-state; see Figure
\ref{fig:time-evolution}. For an inertial observer, the out-state is
inserted precisely at the extra point (the north pole) that $S^{D-1}$
has compared with $\mathbb{R}^{D-1}$. In a stereographic projection of
the sphere to flat Euclidean space, the outgoing state would be at
infinity.

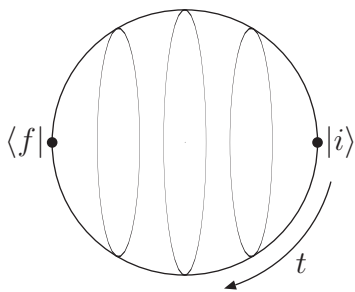
\begin{figure}[hbtp]
\centering
\begin{picture}(0,120)(0,-70)
\BCirc(0,0){50}
\Oval(0,0)(50,8)(0)
\Oval(25,0)(43,8)(0)
\Oval(-25,0)(43,8)(0)
\Vertex(50,0){2}
\Vertex(-50,0){2}
\Text(53,0)[l]{$|i\rangle$}
\Text(-53,0)[r]{$\langle f|$}
\LongArrowArcn(0,0)(57,-15,285)
\Text(42,-42)[tl]{$t$}
\Vertex(0,0){0}
\end{picture}
        \caption{\small Radial quantization on an $S^{D-1}$. In-states
      and out-states are at antipodal points. The Hamiltonian is the
      dilation operator. Each surface corresponding to constant time
      for the observer in the bulk is an $S^{D-2}$.}
      \label{fig:radial-quantization}
\end{figure}

The corresponding situation on the boundary is depicted in Figure
\ref{fig:radial-quantization}. In conclusion, the $\ZZ$ identification
implies that the holographic CFT is simply a theory with conventional
radial quantization on an ordinary sphere. We will see, however, that
the hermiticity conditions of the theory are somewhat unusual.

\subsection{The existence of an S-matrix and holography}
\label{sec:s-matrix}

Defining an S-matrix for quantum gravity in global de Sitter space is
tricky. The problem is that, having defined in- and out-states on two
disconnected surfaces ($\scriminus$ and $\scriplus$), the only
available pairing between them, CPT, is used merely to define an inner
product. Since in quantum gravity the spacetime between these two
boundaries fluctuates, there does not seem to be another way to map
states on $\scriminus$ to $\scriplus$. Hence it is not obviously clear
how to define an S-matrix. If we consider only the quantum field
theory of matter (and neglect back-reaction) with the geometry fixed,
then we are able to define an S-matrix, but even then its matrix
elements are not physically measurable, since no observer can
determine the state at both $\scriminus$ and $\scriplus$, even in the
far future.

In elliptic de Sitter space the situation is different. The past and
future asymptotic regions have been identified, so initial and final
states can be defined in the same asymptotic region, where the
fluctuations of the metric are set to zero. It is useful to think
about the initial and final states in terms of the asymptotic boundary
conditions of various fields, including the metric, in this single
asymptotic region. As discussed in the previous subsection, an
observer positioned at the north pole will use the asymptotic data on
the northern hemisphere to define the in-state and the data on the
southern hemisphere to define the out-state. First, to define an inner
product one can use the canonical map from the north to south pole
which associates to a state $|\Psi_i \ket$ its CPT conjugate state
$\bra \Psi_i|$. Next, to define the S-matrix one uses the combined
asymptotic data provided in the in- and out-states, $|\Psi_i\ket$ and
$\bra \Psi_f|$, as boundary conditions for the ``functional integral''
over all fields in the bulk of the quantum de Sitter space. This
produces a number that can then be identified with the S-matrix
element $\bra \Psi_f|\Psi_i\ket$.

We will now discuss how these S-matrix elements would be possibly
described in a holographic description of de Sitter space. So let us
suppose that elliptic de Sitter space allows a holographic description
in terms of a dual theory, which for concreteness assume to be a
conformal field theory (CFT). Since there is only one asymptotic region
one is dealing with a single euclidean CFT living on a $D\!-\!1$
sphere, which one can think of as the $S^{D-1}$ at $\scriplus$ or
$\scriminus$. In a CFT states can be defined using radial
quantization. They are created by the action of some (local) operator
at the origin:
\begin{equation}
|j \ket = {\cal O}_j (0) | {\rm vac} \ket \; ,
\end{equation}
where we have used the operator-state correspondence. The state $| {\rm
vac} \ket$ is the ``vacuum,'' by which we mean not necessarily a state
of lowest energy (since energy is harder to define in de Sitter
space), but rather a de Sitter-invariant state. Similarly, we can
define a final state as
\begin{equation}
\bra j | = \bra {\rm vac} | {\cal O}^*_j (\infty) \; .
\end{equation}
Notice that this also involves complex conjugation, since our $\ZZ$
map includes charge conjugation, C. Now we can define an inner product
via
\begin{equation}
\bra {\cal O}^*_i (\infty) {\cal O}_j (0) \ket_{S^{D-1}} \equiv
\delta_{ij} \; .
\end{equation}
This pairing of an operator with its CPT conjugate provides an
inner product in the sense of being a map ${\cal H} \times {\cal H}
\to \mathbb{C}$ that is linear in one argument and antilinear in the
other. 

If indeed there is a CFT dual of (elliptic) de Sitter space then,
intuitively, one expects that interactions (and hence S-matrix
elements) are encoded in the correlation functions and/or the operator
product expansion. It is important to note that a CFT by itself does
not have an S-matrix. Therefore instead of studying just the
asymptotic states, let us consider operator insertions at points other
than the origin and infinity. There are an infinite number of such
operators since we can associate an operator to every point on the
sphere. So in principle one could define an infinite set of in-states
by considering strings of operators acting on the in-vacuum,
\begin{equation}
\label{operators}
|\Psi_i\ket = {\cal O}_{j_1}(x_1)\ldots {\cal O}_{j_n} (x_n)
| {\rm vac}\ket \; ,
\end{equation}
and similarly for the out-states. S-matrix elements are then expressed
as correlation functions where part of the operators, those on the
northern hemisphere, represent the in-state, while the other operators
on the southern hemisphere represent the out-state. Note, however,
that not all of these states are independent, because there are
operator product relations. For example, two operators ${\cal O}_i$
and ${\cal O}_j$ inserted at different points have an operator product
relation of the form
\begin{equation}
{\cal O}_i (x_i) {\cal O}_j (x_j) = \sum_k {c^k_{ij} \over |x_i -
x_j|^{\Delta_i+ \Delta_j - \Delta_k}} {\cal O}_k (x_j) \; .
\end{equation}
Here the sum on the right hand side includes (quasi-)primary operators
as well as their descendants. If one allows descendants of arbitrary
conformal dimension, then all operators can be moved to one preferred
point by simply using the Taylor expansion. One natural way to reduce
the redundancy in the states is to consider only quasi-primary
operators. Note that since the conformal dimension of an operator
corresponds to the energy as seen by an observer in de Sitter space,
it is physically reasonable to consider only operators with conformal
dimensions that are below a certain threshold. The number of
(quasi-)primary fields below a certain conformal dimension is
finite. It is natural to conjecture that this fact is related to the
finiteness of the de Sitter entropy. However, note that when one
allows the operators to be inserted at arbitrary points on the sphere,
this still would give an infinite number of states. It may very well
be that there are additional requirements that one has to impose, but
without a more definite and concrete theoretical foundation one can
only guess what these requirements could be.

The most specific proposal that we have for the de Sitter ``S-matrix''
is that it is given by the overlap of the initial and final states
\begin{equation}
S_{fi} = \bra \Psi_f | \Psi_i \ket \; ,
\end{equation}
where both $|\Psi_i\ket$ and $\bra \Psi_f|$ are expressed as in
Eq.~(\ref{operators}) in terms of (quasi-)primary operators with
restricted conformal dimensions. Hence the S-matrix elements are just
given by the correlation functions of the boundary conformal field
theory. This proposal is truly holographic, since the correlation
function are computed in terms of the CFT at the boundary. 

\subsection{Observer complementarity}
\label{sec:complementarity}

How do different observer interpret these S-matrix elements? In fact,
the same operator insertions at the boundary are interpreted
differently by different observers in the bulk. This is because the
physical states defined above depended on a choice of origin. For any
observer, the incoming states are those that correspond to insertions
made on the hemisphere closest to the origin, while outgoing states
are created by operator insertions in the hemisphere nearest to the
antipode of the origin, i.e. at infinity.  Different observers have
different origins so this leads to different interpretations of a
given set of operator insertions. This is observer complementarity.

\begin{figure}[hbtp]
\centering
\begin{picture}(0,180)
\SetScale{1}

\Text(-100,150)[l]{a)}
\BCirc(-50,150){20}
\Line(-50,130)(-50,170)
\LongArrow(-45,150)(-55,150)
\Vertex(-70,150){2}
\Vertex(-32.68,160){2}
\Vertex(-32.68,140){2}

\Photon(30,150)(30,170){1}{4}
\Line(30,150)(45,135)
\Line(30,150)(15,135)
\Text(70,150)[l]{Pair annihilation}

\Text(30,173)[b]{$\gamma$}
\Text(15,133)[t]{$e^-$}
\Text(45,133)[t]{$e^+$}

\Text(-74,150)[r]{$\gamma$}
\Text(-29,162)[l]{$e^+$}
\Text(-29,138)[l]{$e^-$}

\Text(-100,90)[l]{b)}
\BCirc(-50,90){20}
\Line(-50,70)(-50,110)
\LongArrow(-55,90)(-45,90)
\Vertex(-70,90){2}
\Vertex(-32.68,100){2}
\Vertex(-32.68,80){2}

\Photon(30,70)(30,90){1}{4}
\Line(30,90)(45,105)
\Line(30,90)(15,105)
\Text(70,90)[l]{Pair creation}

\Text(30,67)[t]{$\gamma$}
\Text(45,107)[b]{$e^-$}
\Text(15,107)[b]{$e^+$}

\Text(-74,90)[r]{$\gamma$}
\Text(-29,102)[l]{$e^-$}
\Text(-29,78)[l]{$e^+$}

\Text(-100,30)[l]{c)}
\BCirc(-50,30){20}
\Line(-68.794,23.16)(-31.206,36.84)
\LongArrow(-48.29,25.3)(-51.71,34.7)
\Vertex(-70,30){2}
\Vertex(-32.68,40){2}
\Vertex(-32.68,20){2}

\Photon(30,30)(50,30){1}{4}
\Line(15,45)(30,30)
\Line(15,15)(30,30)
\Text(70,30)[l]{Photon emission}

\Text(51,30)[l]{$\gamma$}
\Text(13,45)[r]{$e^-$}
\Text(13,15)[r]{$e^-$}

\Text(-74,30)[r]{$\gamma$}
\Text(-29,42)[l]{$e^-$}
\Text(-29,18)[l]{$e^-$}

\end{picture}
\caption{\small Complementarity in action: the same correlation function as
interpreted by an observer a) at the south pole, b) at the north pole,
and c) at an intermediate point. The circle denotes the sphere on
which the dual theory lives, the dots are operator insertions, the
arrow indicates the observer's direction of time, and the equator
divides the in-states from the out-states. On the right are the
corresponding processes in spacetime.}\label{fig:feynman}
\end{figure}
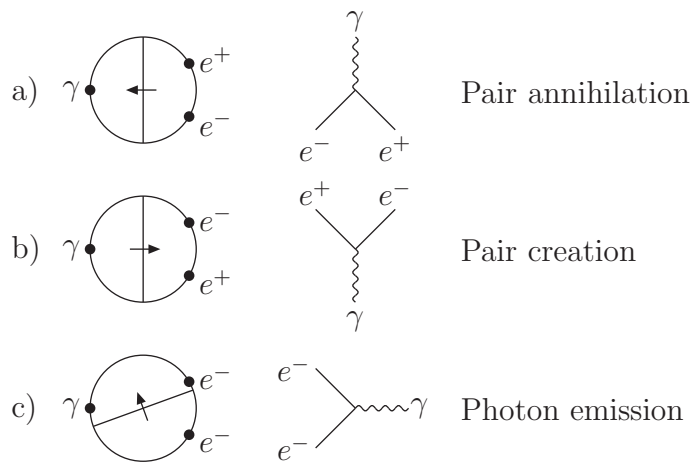

Consider, for example, the situation indicated in Figure
\ref{fig:feynman}. A south pole observer would describe this as pair
annihilation: an electron and a positron come in, and annihilate to
give a photon. On the other hand, a north pole observer, being
antipodal to the south pole observer, would see the same events
happening in a CPT mirror. In this case, it would describe the
CPT-conjugate process of pair creation: an incoming photon decays into
an electron and a positron. A different observer in between these two
poles would see yet another situation, for example, an incoming
electron emitting a photon. All these processes have the same
amplitude.

\subsection{A little group theory}
\label{sec:representation}

A striking consequence of the preceding discussion is that, although
the S-matrix itself is de Sitter-invariant, the in-states themselves
are not. De Sitter transformations that take one observer into another
generically transform in-states into out-states and vice versa. Hence
the asymptotic Hilbert space does not decompose into irreducible
representations of the de Sitter group. This is important because
there is a well-known theorem which states that (nontrivial) unitary
representations of noncompact groups must be
infinite-dimensional. This theorem is in tension with the finiteness
of the de Sitter entropy. If the de Sitter entropy enumerates the
microscopic degrees of freedom underlying a quantum description of de
Sitter space, then we would expect it to form a (possibly reducible)
representation of some group. Were that group to be the noncompact de
Sitter group, $O(1,D)$, then the holographic theory could not be
unitary. For elliptic de Sitter space, the entropy is presumably also
given by the Bekenstein-Hawking formula:
\begin{equation}
S = {A \over 4} = {\pi^{D-1 \over 2} R^{D-2} \over 4 \Gamma \left( {D-1
\over 2} \right)} \; ,
\end{equation}
where the ``area,'' $A$, is the volume of the horizon which is now a
\dimmin{D}{2} real projective sphere, $\mathbb{R}P^{D-2}$. The
important point here is that this is again finite. But as we saw, the
states in elliptic de Sitter space do not transform under
representations of the full de Sitter group. Instead, they only
transform under the subgroup that preserves the asymptotic position of
an observer. Since asymptotically an observer is a point on a
\dimmin{D}{1} sphere (and in the future, a possibly different point on
the same sphere), the relevant group is actually $SO(D-1)$. We propose
that the entropy of de Sitter space is related to representations of
this compact group.

Another way to make the same point is as follows. The
Bekenstein-Hawking entropy refers to the area of a holographic screen
bounding a given region of spacetime. For de Sitter space, a horizon
is actually the holographic screen of a particular observer in the far
future. But the screen accessible to any single observer must furnish a
representation of the {\em little group} of that observer. This is
precisely the rotation group, $SO(D-1)$.

A given physical state is therefore labeled by its conformal weight,
its angular momenta, and the quantum numbers of any internal
symmetries. Nevertheless it is still a great challenge to show that
the number of such states is precisely $\exp(A/4)$. In principle, the
conformal weights and angular momenta could be arbitrarily high,
leading to representations that would be too big. One possibility
might be to restrict the maximum scaling dimension
\begin{equation}
\Delta_i \leq \Delta_{\rm max}
\end{equation}
of any state $|i \ket$. Here the idea is that the scaling weight is
the eigenvalue of the CFT Hamiltonian, but we know that energy in de
Sitter space is bounded by the mass of the largest black hole that can
fit within the de Sitter horizon. This suggests that we should only
consider those states which have scaling dimension below some maximum.

\subsection{Hermiticity}
\label{sec:hermiticity}

It is usually accepted that the holographic dual to de Sitter space
must be a nonunitary theory. The argument considers fields propagating
in the bulk spacetime. We can take the field to be a massive scalar
field; higher-spin fields are qualitatively similar. In planar
coordinates valid near \scriminus, the line element is
\begin{equation}
ds^2 = -dt^2 + e^{-2t/R} dx^2_d \; ,
\end{equation}
and the scalar wave equation is
\begin{equation}
- \partial^2_t \phi + {d \over R} \partial_t \phi + e^{2t/R} \nabla^2
\phi - m^2 \phi = 0 \; .
\end{equation}
Near \scriminus, as $t \to -\infty$ the field asymptotically behaves
like $\phi(t,x) \sim e^{h_+ t/R} f(x) + e^{h_- t/R} g(x)$, where
\begin{equation}
h_\pm = {1 \over 2} \left( d \pm \sqrt{d^2 - 4m^2 R^2} \right) \; .
\end{equation}
Notice that for sufficiently high mass this is complex. In terms of
the boundary theory, there seem to be operators with complex scaling
dimension in the CFT. This would suggest that the theory contains
states of negative norm. Let us review the reasoning that leads to this
conclusion.

Consider, at first, three-dimensional de Sitter space. The conformal
field theory lives on a two-sphere, or the complex plane.
Recall that with radial quantization on the complex plane, the in- and
out-states are related by BPZ conjugation, a purely analytic (or
purely antianalytic) map:
\begin{equation}
z \to -1 /z \; .
\end{equation}
The BPZ map takes the origin to complex infinity while preserving the
upper half plane, allowing us to define a relation between bras and kets:
\begin{equation}
|\phi \ket = \phi(0,0) |0 \ket \to \bra 0| \phi(\infty,
\infty) = \bra \phi | \equiv | \phi \ket^\dagger \; .
\end{equation}
In other words, the BPZ map motivates the usual choice of Hermitean
conjugation for the Virasoro generators:
\begin{equation}
L^{\dagger}_n = L_{-n} \qquad \cc{L}^{\dagger}_n = \cc{L}_{-n} \; .
\end{equation}
A direct consequence of this is that primary fields with complex conformal
weights lead to descendants with complex norm:
\begin{equation}
|| L_{-1} |h \ket ||^2 = \bra h | L_1 L_{-1} |h \ket = 2 h
\bra h | h \ket \; .
\end{equation}
Thus a sufficiently massive scalar field in de Sitter space seems to
lead to a nonunitary conformal field theory.

Now consider the antipodal identification. We can express the line
element in global coordinates as
\begin{equation}
ds^2 = -dt^2 + 4 R^2 \cosh^2 (t/R) {dz d\cc{z} \over (1 + |z|^2)^2} \; .
\end{equation}
The antipodal map is
\begin{equation}
t \to -t \qquad z \to -1/\cc{z} \qquad \cc{z} \to -1/z \; .
\end{equation}
Holomorphic and antiholomorphic coordinates are interchanged!
Incoming states created by holomorphic fields at $t = - \infty$ are
taken to antiholomorphic final states at $t = + \infty$, and vice
versa. Hence
\begin{equation}
L^\dagger_n = \cc{L}_{-n} \qquad \cc{L}^\dagger_n = L_{-n} \; . \label{dagger}
\end{equation}
With this definition of Hermitean conjugation, certain states with
complex conformal weights now have positive norm. Such a hermiticity
condition has also been proposed in \cite{janvijay}. Consider a primary
field with complex conjugate weights $h_{\pm}$. Acting on the
corresponding state with $L_{-n} \cc{L}_{_n}$ gives a state of the
form $|\phi \ket = L_{-n} \cc{L}_{-n}$. Its norm is
\begin{equation}
\bra \phi | \phi \ket = \left( 4 n^2 |h|^2 + {c^2 \over 144} (n^3 -
n)^2 + {c \over 6} (n^4 - n^2) (h + \cc{h}) \right) \bra h, \cc{h}
\hws \; ,
\end{equation}
which is real and positive, even though $h$ may be complex. The rule
is that to have positive norm, the total level of $L$ and $\cc{L}$
must be the same. States for which the levels of $L$ and $\cc{L}$ do
not match have zero norm. These include states like $L_{-1} \hws$,
which would have had positive norm (for real $h, \cc{h}$) with the
conventional definition of Hermitean conjugation. However, linear
combinations of zero norm states can still lead to states of negative
norm. So there is still the danger that the dual CFT is nonunitary.

We note, however, that nonunitarity in the spectrum of descendants of
the CFT may not necessarily be a problem for its use as a dual for
elliptic de Sitter space. This is because, as we discussed above,
states that have a physical meaning in this context may have to
satisfy additional requirements, such as that they are
quasi-primary. In this case, states like $L_{-n} \cc{L}_{-n} \hws$ are
not physical states. For example, the fact that $L_{-1}$ acting on a
physical state does not lead to a physical state could be a
consequence of the fact that translations of the entire state of the
universe are not represented in the Hilbert space of a single
observer, since such translations also change the location of the
observer. If one considers only highest weight states (those created
by quasi-primary operators acting on the vacuum), then there is no
problem of negative norm states. Note that restriction to highest
weight states reduces the number of states: it effectively subtracts
one from the total central charge. But since we expect $c \gg 1$ this
does not change the counting of states significantly.

The generalization of this discussion to higher dimensions is
straightforward. Writing the de Sitter line element as
\begin{equation}
ds^2 = -dt^2 + 4R^2 \cosh^2 (t/R) {dx^2 \over ( 1 + r^2 )^2 } \; ,
\end{equation}
where $r^2 = |\vec{x}|^2$, the antipodal map takes
\begin{equation}
t \to -t \qquad  x^i \to -{x^i \over r^2} \; .
\end{equation}
The conformal generators in higher dimensions are $D$, the dilatation
operator, $K_a$, the special conformal transformations, as well as the
rotations, $J_{ab}$, and the translations, $P_a$. The antipodal map
suggests that the hermiticity properties should be
\begin{equation}
D^\dagger = D \qquad J^\dagger_{ab} = J_{ab} \qquad P^\dagger_a = K_a
\qquad K^\dagger_a = P_a \; .
\end{equation}
Once again, the translations and special conformal transformations do
not preserve the set of physical states. The physical states are
labeled by the hermition operators which are labeled by the
simultaneous eigenvalues of $D$, $J_{ab}$, and a Cartan set of any
internal symmetry group.

\section{On a String Realization}
\label{sec:string-realization}

Our discussion of the elliptic interpretation of de Sitter space and
its holographic implementation has been rather intuitive. Clearly, to
make things more precise one needs a concrete realization of these
ideas in a working theory of quantum gravity, such as string theory
(or perhaps loop gravity \cite{smolin}). It has been surprisingly hard
to find a realization of de Sitter space in string theory. One
obstacle to a satisfactory string-theoretic description of de Sitter
space is the lack of supersymmetry. Intuitively, de Sitter space
cannot be supersymmetric because it is thermal; at finite temperature
bosons and fermions have different statistics. More formally, there is
no superalgebra that contains the de Sitter isometry group and is
represented by Hermitean supercharges. The known super-extensions of
the de Sitter isometry group \cite{Newhouses} involve nonpositive
quadratic forms and have no unitary representations. This difficulty
can be traced back to the fact that there is no globally-defined
timelike Killing vector in de Sitter space, and hence there is no
positive-definite Hamiltonian, $H$. This same non-positive-definite
nature shows up in attempts to construct de Sitter space using
timelike T-duality and/or compactifications on noncompact Euclidean
manifolds \cite{Hull98,Hull01}. The resulting gauged supergravity
theories allow de Sitter space as a solution but have ghosts,
i.e. fields with kinetic terms of the wrong sign.

The nature of these problems changes in elliptic de Sitter space,
mainly because it is not a time-orientable space. In fact, we would
like to believe that the only possible realization of de Sitter space
in string theory is in its elliptic form. The failure to find a de
Sitter solution in string theory may well be that one should perhaps
have been looking at string backgrounds that are not
time-orientable. Clearly, time-unorientability poses new challenges
for string theory, and it is not immediately obvious how it can be
defined consistently \cite{vijayesko}. In this respect, it is
interesting that de Sitter space arises in type IIB* string theory
after a timelike T-duality, which can be thought of as a change of
sign of the left- (or right-) moving part of the worldsheet scalar
$X^0$ corresponding to time. Hence, after a T-duality it is as if the
right- (or left-) movers go forward in time, while the left- (or
right-) movers go backward in time. Perhaps this means that type IIB*
string theory has to be quantized in a different way so that
worldsheets and/or the spacetime background have to be
time-unorientable. This may change the problems with ghost-like fields
and perhaps solve it. We hope to report on this issue in the future.

Now let us make some observations on the candidate conformal field
theory dual of five-dimensional elliptic de Sitter space as suggested
by its realization in IIB* string theory. Type IIB* theory can be
thought of as arising through a timelike T-duality of type IIA theory
\cite{Hull98,Hull01}. The low-energy limit of IIB* theory is IIB*
supergravity which has Dirichlet brane solutions that have purely
spatial extent; they are called E$p$-branes when their worldvolume is
$p$-dimensional. Following Hull we consider the near-horizon geometry
of a stack of $N$ E4-branes, which are the Euclidean analogs of the
D3-branes of type IIB theory. The metric resembles that of the
D3-brane,
\begin{equation}
ds^2 = H^{-1/2}(\rho) dx_\parallel^2 + H^{1/2} (\rho) dx_\perp^2 \; ,
\end{equation}
where $H(\rho)$ is the usual harmonic function,
\begin{equation}
H(\rho) = 1 + {4 \pi {\alpha^\prime}^2 g N \over \rho^4} \; ,
\end{equation}
except that, because the branes are Euclidean, the transverse
``radius'' also includes time:
\begin{equation}
\rho^2 = x_{\perp}^2 = \vec{x}^2 - t^2 \; .
\end{equation}
The horizon is at $\rho = 0$. Now we would like to take the
near-horizon limit. Since $\rho$ depends on time, there are two ways
we can approach the horizon, where $\rho$ is timelike and where
$\rho$ is spacelike. For spacelike $\rho$ the transverse geometry is
\begin{equation}
dx_\perp^2= -dt^2 + d\vec{x}^2 = d\rho^2 + \rho^2 ds_{dS_5}^2 \; ,
\end{equation}
where $ds_{dS_5}^2$ is the line element of five-dimensional de Sitter
space. For timelike $\rho$ we get instead
\begin{equation}
dx_\perp^2 = -d\rho^2 + \rho^2 ds_{H^5}^2 \; ,
\end{equation}
where $H^5$ is the five-dimensional hyperbolic (Lobachevsky) plane
(i.e. Euclidean anti-de Sitter space). In the near-horizon
limit we drop the 1 in $H(\rho)$ to obtain, for spacelike $\rho$,
\begin{equation}
ds^2 = \lf \sqrt{4 \pi {\alpha^\prime}^2 g N} ~ { d \rho^2 \over \rho^2} +
{\rho^2 \over \sqrt{4 \pi {\alpha^\prime}^2 g N} } dx_\parallel^2 \rt
+ \sqrt{4 \pi {\alpha^\prime}^2 g N} ~ ds_{dS_5}^2 \; .
\end{equation}
The geometry is therefore locally that of $H^5 \times dS_5$. For
timelike $\rho$ we obtain
\begin{equation}
ds^2 = \lf \sqrt{4 \pi {\alpha^\prime}^2 g N} ~ { - d \rho^2 \over \rho^2} +
{\rho^2 \over \sqrt{4 \pi {\alpha^\prime}^2 g N} } dx_\parallel^2 \rt
+ \sqrt{4 \pi {\alpha^\prime}^2 g N} ~ ds_{H^5}^2 \; . \label{tlrho}
\end{equation}
This too is $dS_5 \times H^5$. So again we get the same local
geometry. However, there are some important differences between the
two. For spacelike $\rho$, the branes are part of $H^5$, and de Sitter
space is part of the transverse space; that is not what we want. For
timelike $\rho$, the branes are part of de Sitter space and $H^5$ is
transverse. So we should choose $\rho$ to be timelike. The E4-branes
are now on the boundary of de Sitter space, at $\scri$. But now note
that there are two disconnected branches because in foliating
Minkowski space into spacelike slices (which corresponds to timelike
$\rho$) one can have $t > 0$ or $t < 0$. In order to have a connected
geometry, we should really identify these two branches by making a
$\mathbb{Z}_2$ identification. In that case the metric that we just
described must be modded out by a $\mathbb{Z}_2$ that maps $t\to -t$.
Since the line element on de Sitter space in Eq.~(\ref{tlrho}) covers
one inflationary patch, an identification of $t$ and $-t$ suggests
that the near-horizon geometry becomes $edS_5 \times H^5$. A
$\mathbb{Z}_2$ identification of the transverse geometry implies that
the E4-branes are on a $T$-orientifold, the purely spatial counterpart
of a conventional orientifold. Indeed, elliptic de Sitter space is the
analytic continuation of the $\mathbb{R}P^5$ that arises (instead of
an $S^5$) in the transverse geometry of D3-branes on an orientifold
plane.

The theory on the worldvolume of the E4-brane is Euclidean ${\cal N}
=4$ SYM.  This theory is obtained from ${\cal N} = 1$ SYM in
$D=9\!+\!1$ by dimensional reduction, where one of the
compactification directions is time. So one of the six scalars in the
E4 worldvolume theory comes from the timelike component of the
\dimplus{9}{1} gauge field. This becomes a scalar with the wrong sign
kinetic operator, and therefore we are dealing with a conformal field
theory with a ghost. In fact, there are several reasons to expect such
ghost fields to be present in a CFT dual to de Sitter space. First,
the six scalars form a vector $(\phi_0,\vec{\phi})$ of the $SO(1,5)$
R-symmetry of the Euclidean theory; invariance under the R-symmetry
already implies that one scalar has the wrong sign kinetic term. A
second reason is the following.

Just as in AdS/CFT one expects the holographic direction to correspond
to the RG-scale of the dual field theory. But unlike in AdS/CFT, the
holographic direction is timelike in de Sitter space. This timelike
nature of the RG-scale is directly related to the presence of the
ghost scalar. Namely, the energy scale $\mu$ of the theory can be
defined in terms of the values of the scalar fields as
\begin{equation}
\langle \vec{\phi}^2-\phi_0^2\rangle =\pm\mu^2 \; .
\end{equation}
Let us now fix the energy scale $\mu$. The scalar fields are then
restricted to a five-dimensional scalar manifold. Here we have a
choice: for the $-$ sign the resulting scalar manifold is the
Lobachevsky plane, while for the $+$ sign it is de Sitter space. If we
take the $+$ sign the $\phi_0$ field still has fluctuations with the
wrong sign. However, if we take the $-$ sign, all the fluctuations of
the scalar field have the correct sign in their kinetic term.

The parameter $\mu$ becomes the renormalization group scale, and in
fact is the same as the holographic time coordinate $\rho$: together
with the four Euclidean coordinates on the E4-brane, it leads to de
Sitter space. As we noted, the scalar manifold has two disconnected
branches, corresponding to $\phi_0 >0$ and $\phi_0 < 0$. Now here
there is a difference between $U(N)$ and $SO(N)$ SYM. In the latter
case one can use the gauge symmetry to map $\phi$ to $-\phi$. This
identifies the two branches of the scalar manifold. An $SO(N)$ gauge
group arises if we put $N$ coincident E4-branes on top of a
$T$-orientifold plane. This is precisely what we argued for
earlier. In the near-horizon limit we get antipodally-identified de
Sitter space. So finally, we come to the following conjecture: the
large-N limit of $SO(N)$ SYM theory, with conformal group $SO(1,5)$
and R-symmetry group $SO(1,5)$, in the phase described by the $-$ sign
in the scalar equation is the holographic dual of $edS_5 \times H^5$,
or elliptic de Sitter space times a hyperbolic five-plane. There is
now only one boundary, an $S^4$, and that is the boundary on which the
CFT lives.

\section{Conclusion}

In this paper we studied de Sitter space in its elliptic
interpretation with antipodal points identified. We discussed several
conceptual issues in the context of the elliptic interpretation,
especially questions regarding holography and the definition of an
S-matrix. Our conclusions support the view that the antipodal
identification does make sense and in fact may even be required to
arrive at a consistent description of de Sitter quantum gravity. The
presented arguments in favor of the antipodal identification range
from suggestive to rather compelling; they are not yet sufficient to
claim that antipodal identification is the only way to view quantum de
Sitter space.

{}From our point of view the most convincing arguments supporting the
elliptic de Sitter space are: a) the implementation of observer
complementarity: all observers have complete information, but have
different interpretations, and b) the realization of holography: for
every observer time-evolution and the S-matrix are naturally described
in terms of a dual theory on a single boundary. The most serious
challenge to elliptic de Sitter space is the issue of possible closed
timelike curves after including backreaction. Once gravitational
backreaction is taken into account, the Penrose diagram of perturbed
de Sitter space becomes a ``tall'' rectangle
\cite{GaoWald,talltales}. This implies that certain antipodal points
come into causal contact. The resulting closed timelike curves are
contained in the bulk of de Sitter space, and therefore it is not
immediately obvious how it would affect the theory on the
boundary. One point of view is that the perturbation of de Sitter
space should be described by an appropriately perturbed CFT, for which
the holographic reconstruction breaks down at some point in the
bulk. The presciptions for the time evolution of a single observer and
for his observable S-matrix are, however, defined purely in terms of
the boundary and could still make sense. Clearly this issue needs
further study.

Finally, the most pressing open issue is whether one can find a
consistent description of de Sitter space in string theory, or perhaps
in some other working theory of quantum gravity. There are many
reasons to believe that such a description would be holographic and
will incorporate a version of observer complementarity. We are hopeful
that the ideas presented in this paper will then in some form be fully
realized.

\bigskip
\noindent
{\bf Acknowledgments}

We have benefited from conversations with Jan de Boer, Raphael Bousso,
Chang Chan, Brian Greene, Chris Hull, Dan Kabat, Finn Larsen, Sameer Murthy,
Rob Myers, Joe Polchinski, Koenraad Schalm, Lee Smolin, Marcus Spradlin, and
Leonard Susskind.

\noindent {M.~P.} is supported by DOE grant DF-FCO2-94ER40818. {I.~S.} is
supported by the Dutch Science Foundation (NWO). {E.~V.} is supported by DOE
grant DE-FG02-91ER40571.

\end{document}